\documentstyle[preprint,aps,eqsecnum]{revtex}
\begin{document}
\draft
\tighten

\title{Qutrit Entanglement\thanks{We dedicate this article to Marlan
Scully and his lifelong interest in the weird and wonderful world of 
quantum mechanics.  We understand that the article will appear in the 
Festschrift celebrating Marlan's 60th birthday, but surely that 60 is 
a mistake.  Perhaps Marlan's passport says he's 60, but that's because 
his passport doesn't know him.  He approaches physics with the same 
curiosity and zest he displayed when we met him nearly twenty years ago.  
Happy 60th, Marlan!  We draw inspiration from your dedication to and 
enthusiasm for physics.}} 

\author{Carlton M.~Caves$^{\rm (a)}$ and Gerard J.~Milburn$^{\rm(b)}$}

\address{
$^{\rm (a)}$Center for Advanced Studies, Department of Physics and Astronomy,\\
University of New Mexico, Albuquerque, New Mexico 87131-1156, USA\\
$^{\rm (b)}$Centre for Laser Science, Department of Physics,\\
The University of Queensland, St Lucia, Queensland 4072, Australia}

\date{\today}

\maketitle

\begin{abstract}
We consider the separability of various joint states for $N$ qutrits. 
We derive two results: (i) the separability condition for a two-qutrit state 
that is a mixture of the maximally mixed state and a maximally entangled 
state (such a state is a generalization of the Werner state for two qubits);
(ii) upper and lower bounds on the size of the neighborhood of separable
states surrounding the maximally mixed state for $N$ qutrits.
\end{abstract}
\pacs{}

\section{Introduction}
\label{sec:intro}

Quantum entanglement provides a powerful physical resource for new kinds 
of communication protocols and computation \cite{Lo1998}, which achieve 
results that cannot be achieved classically.  Quantum entanglement refers 
to correlations between the results of measurements made on component
subsystems of a larger physical system that cannot be explained in terms 
of correlations between local classical properties inherent in those same 
subsystems.  Alternatively, an entangled state cannot be prepared by local 
operations and local measurements on each subsystem.  While the nonclassical
nature of quantum entanglement was recognized by Einstein, Schr{\"o}dinger, 
and other pioneers of quantum mechanics, it is only recently that a full
appreciation of the surprising and complex properties of entanglement has 
begun to emerge. 

We now have a good understanding of entanglement for a pair of qubits
\cite{Wootters1998}, a qubit being a system with a two-dimensional Hilbert 
space.  Moreover, we have a criterion, the partial transposition condition 
of Peres \cite{Peres1996}, which determines whether a general state of
two qubits is entangled and whether a general state of a qubit and a qutrit
is entangled \cite{MHorodecki1996}, a qutrit being a system whose states
live in a three-dimensional Hilbert space.  The partial-transposition 
condition fails, however, to provide a criterion for entanglement in other 
cases, where the constituents have higher Hilbert-space dimensions 
\cite{PHorodecki1997,Lewenstein1999} or where there are more than two
constituents.  In this paper we consider the entanglement of composite 
systems made of qutrits.  In particular, we investigate the separability 
of various joint states of two or more qutrits.  A joint state of a composite 
system is defined to be {\it separable\/} if it can be decomposed into a 
mixture of product states for the constituents.  A separable state has no 
quantum entanglement.

In Sec.~\ref{sec:qutrits}, we review an efficient operator representation
of the states of a qutrit \cite{Arvind1997}, particularly the pure states. 
This operator representation is analogous to the familiar Pauli or 
Bloch-sphere representation of qubit states.  The representation is applied
to an analysis of qutrit entanglement in Secs.~\ref{sec:epmax} and 
\ref{sec:sep}. \ In Sec.~\ref{sec:epmax}, we consider states of two qutrits
that are a mixture of the maximally mixed state and a maximally entangled
state.  Such states are a generalization of the Werner state for two qubits 
\cite{Werner1989}.  We show that the two-qutrit mixture is separable if 
and only if the probability for the maximally entangled state does not 
exceed 1/4.  This result is a special case of a general result obtained
by Vidal and Tarrach \cite{Vidal1999}, which gives, for any bipartite
system, the separability boundary for any mixture of the maximally mixed
state with a pure state of the bipartite system.  In Sec.~\ref{sec:sep}, 
we consider the separability of $N$-qutrit states near the maximally mixed 
state.  We find both lower and upper bounds on the size of the neighborhood 
of separable states around the maximally mixed state.  Our analysis follows 
closely that of Braunstein {\it et al.}\ \cite{Braunstein1999}, who analyzed 
the separability of $N$-qubit states near the maximally mixed state.  

\section{Operator representations of qutrit states}
\label{sec:qutrits}

In this section we review an efficient operator representation for qutrit
states, analogous to the Pauli or Bloch-sphere representation for qubits.  
The qutrit representation uses the (Hermitian) generators of SU(3) as an 
operator basis.  Our discussion is taken from Ref.~\cite{Arvind1997}; the 
reader is referred there for more details (see also Refs.~\cite{Khanna1997} 
and \cite{Byrd1998}).  This operator representation---and its generalization 
to higher dimensions---has been used by Ac{\'i}n, Latorre, and Pascual 
\cite{Acin1999} to explore optimal generalized measurements on collections 
of identically prepared spin systems.

Let $|1\rangle$, $|2\rangle$, and $|3\rangle$ be an orthonormal basis for
a qutrit.  To denote these states, we use Latin letters from the beginning of 
the alphabet, which take on the values 1, 2, and 3.  A pure state 
$|\psi\rangle$ is a superposition of the states $|a\rangle$.  Normalization 
and a choice for the arbitrary overall phase allow us to write any pure state 
as
\begin{equation}
|\psi\rangle=
e^{i\chi_1}\sin\theta\cos\phi\,|1\rangle
+e^{i\chi_2}\sin\theta\sin\phi\,|2\rangle
+\cos\theta\,|3\rangle
\;.
\label{eq:psi}
\end{equation}
One obtains all pure states as the four co\"ordinates vary over the ranges
\begin{mathletters}
\begin{equation}
0 \leq \theta,\phi \leq \pi/2\;,
\end{equation}
\begin{equation}
0 \leq \chi_1,\chi_2 \leq 2\pi\;.
\end{equation}
\label{eq:ranges}
\end{mathletters}

To develop an operator representation of qutrit states, we begin with 
the eight (Hermitian) generators of SU(3).  In the basis $|a\rangle$, 
these generators have the matrix representations
\begin{eqnarray}
\lambda_1&=&\pmatrix{0&1&0\cr 1&0&0\cr 0&0&0}\;,\qquad
\lambda_2=\pmatrix{0&-i&0\cr i&0&0\cr 0&0&0}\;,\qquad
\lambda_3=\pmatrix{1&0&0\cr 0&-1&0\cr 0&0&0}\;,\nonumber\\
\lambda_4&=&\pmatrix{0&0&1\cr 0&0&0\cr 1&0&0}\;,\qquad
\lambda_5=\pmatrix{0&0&-i\cr 0&0&0\cr i&0&0}\;,\label{eq:generators}\\
\lambda_6&=&\pmatrix{0&0&0\cr 0&0&1\cr 0&1&0}\;,\qquad
\lambda_7=\pmatrix{0&0&0\cr 0&0&-i\cr 0&i&0}\;,\qquad
\lambda_8={1\over\sqrt3}\pmatrix{1&0&0\cr 0&1&0\cr 0&0&-2}\;.\nonumber
\end{eqnarray}
We use Latin indices from the middle of the alphabet, ranging from 
1 to 8, to label these generators and related quantities, and we use
the summation convention to indicate a sum on repeated indices.  The 
generators~(\ref{eq:generators}) are traceless and satisfy
\begin{equation}
\lambda_j\lambda_k=
{2\over3}\,\delta_{jk}
+d_{jkl}\lambda_l
+if_{jkl}\lambda_l
\;.
\label{eq:genprod}
\end{equation}
The coefficients $f_{jkl}$, the structure constants of the Lie group SU(3), are 
given by the commutators of the generators and are completely antisymmetric 
in the three indices.  The coefficients $d_{jkl}$ are given by the 
anti-commutators of the generators and are completely symmetric.  Values 
of these coefficients can be found in Ref.~\cite{Arvind1997}.  

By supplementing the eight generators with the operator
\begin{equation}
\lambda_0\equiv\sqrt{2\over3}\,1\;,
\end{equation}
we obtain a Hermitian operator basis for the space of linear operators
in the qutrit Hilbert space.  This basis is an orthogonal basis, satisfying
\begin{equation}
{\rm tr}(\lambda_\alpha\lambda_\beta)=2\delta_{\alpha\beta}\;.
\end{equation}
Here and throughout Greek indices run over the values 0 to 8.

Any density operator can be expanded uniquely as
\begin{equation}
\rho={1\over3}c_\alpha\lambda_\alpha\;,
\label{eq:rhooneexp}
\end{equation}
where the (real) expansion coefficients are given by
\begin{equation}
c_\alpha={3\over2}{\rm tr}(\rho\lambda_\alpha)\;.
\label{eq:cone}
\end{equation}
Normalization implies that $c_0=\sqrt{3/2}$, so the density operator takes
the form
\begin{equation}
\rho={1\over3}(1+c_j\lambda_j)=
{1\over3}(1+{\bf c}\cdot\bbox{\lambda})
\;.
\end{equation}
Here ${\bf c}=c_j{\bf e}_j$ can be regarded as a vector in a real,
eight-dimensional vector space, and $\bbox{\lambda}=\lambda_j{\bf e}_j$ 
is an operator vector.  Using Eq.~(\ref{eq:genprod}), one finds that
\begin{equation}
\rho^2=
{1\over9}\!\left(1+{2\over3}\,{\bf c}\cdot{\bf c}\right)1+
{1\over3}\,\bbox{\lambda}\cdot
\!\left({2\over3}\,{\bf c}+{1\over3\sqrt3}\,{\bf c}\star{\bf c}\right)
\;,
\end{equation}
where the ``star'' product is defined by
\begin{equation}
{\bf c}\star{\bf d}\equiv{\bf e}_j d_{jkl}c_kd_l
\;.
\end{equation}

For a pure state, $\rho^2=\rho$, so we must have ${\bf c}\cdot{\bf c}=3$ and 
${\bf c}\star{\bf c}=\sqrt3\,{\bf c}$.  Defining the eight-dimensional unit 
vector ${\bf n}={\bf c}/\sqrt3$, we find that any pure state of a qutrit 
can be written as
\begin{equation}
\rho=|\psi\rangle\langle\psi|=
\frac{1}{3}(I+\sqrt{3}\,{\bf n}\cdot\bbox{\lambda})
\equiv P_{\bf n}
\;,
\label{eq:purerho}
\end{equation}
where ${\bf n}$ satisfies
\begin{mathletters}
\begin{eqnarray}
&\mbox{}&{\bf n}\cdot{\bf n}=1\;,\\
&\mbox{}&{\bf n}\star{\bf n}={\bf n}
\label{eq:starcond}
\;.  
\end{eqnarray}
\end{mathletters}
We introduce the notation $P_{\bf n}$ for a pure state with unit vector
${\bf n}$ for use later in the paper.

Equation~(\ref{eq:cone}) implies that 
\begin{equation}
n_j={\sqrt3\over2}{\rm tr}(\rho\lambda_j)=
{\sqrt3\over2}\langle\psi|\lambda_j|\psi\rangle
\;.
\end{equation}
Applied to the pure state~(\ref{eq:psi}), this gives a unit vector
\begin{eqnarray}
{\bf n} & = & \sqrt3 \Biggl(
\sin^2\!\theta\sin\phi\cos\phi\cos(\chi_2-\chi_1)\,{\bf e}_1
+\sin^2\!\theta\sin\phi\cos\phi\sin(\chi_2-\chi_1)\,{\bf e}_2
\nonumber\\
&\mbox{}&\phantom{\sqrt3\Biggl(\;}
+{1\over2}\sin^2\!\theta(\cos^2\!\phi-\sin^2\!\phi)\,{\bf e}_3
+\sin\theta\cos\theta\cos\phi\cos\chi_1\,{\bf e}_4
\nonumber\\
&\mbox{}&\phantom{\sqrt3\Biggl(\;}
-\sin\theta\cos\theta\cos\phi\sin\chi_1\,{\bf e}_5
+\sin\theta\cos\theta\sin\phi\cos\chi_2\,{\bf e}_6
\nonumber\\
&\mbox{}&\phantom{\sqrt3\Biggl(\;}
-\sin\theta\cos\theta\sin\phi\sin\chi_2\,{\bf e}_7
+{1\over2\sqrt3}(1-3\cos^2\!\theta)\,{\bf e}_8 
\Biggr)\;.
\end{eqnarray}

The pure qutrit states lie on the unit sphere in the eight-dimensional
vector space, but not all operators on the unit sphere are pure states.
For example, of the unit vectors $\pm{\bf e}_j$, $j=1,\ldots,8$, only 
$-{\bf e}_8$ satisfies the star-product condition~(\ref{eq:starcond}) 
and thus is the unit vector for a pure state.  It is easy to show that
the unit vectors that do not satisfy the star-product condition do not
specify any state, pure or mixed, for they all give operators that have
negative eigenvalues.  The star-product condition~(\ref{eq:starcond}) 
places three constraints on the unit vector for a pure state, thus reducing 
the number of real parameters required to specify a pure state from the 
seven parameters needed to specify an arbitrary eight-dimensional unit 
vector to four parameters, which can be taken to be the four co\"ordinates 
of Eq.~(\ref{eq:psi}).  

It is useful to notice that
\begin{equation}
|\langle\psi|\psi'\rangle|^2={\rm tr}(\rho\rho')=
{1\over3}(1+2{\bf n}\cdot{\bf n}')\;.
\end{equation}
Orthogonal pure states have unit vectors that satisfy 
${\bf n}\cdot{\bf n}'=-{1\over2}$ and are thus $120^\circ$ apart.  The states 
in an orthonormal basis have unit vectors that lie in a plane at the vertices 
of an equilateral triangle.  The density operators that are diagonal in the 
orthonormal basis are the operators on the triangle or in its interior.

The orthonormal states $|a\rangle$, $a=1,2,3$, of the original basis have 
unit vectors ${\bf n}_a$ that lie in the plane spanned by ${\bf e}_3$ and 
${\bf e}_8$: 
\begin{mathletters}
\begin{eqnarray}
{\bf n}_1&=&{\sqrt3\over2}{\bf e}_3+{1\over2}\,{\bf e}_8
\quad\mbox{($\theta=\pi/2$, $\phi=0$, $\chi_1$ and $\chi_2$ arbitrary),}\\
{\bf n}_2&=&-{\sqrt3\over2}{\bf e}_3+{1\over2}\,{\bf e}_8
\quad\mbox{($\theta=\pi/2$, $\phi=\pi/2$, $\chi_1$ and $\chi_2$ arbitrary),}\\
{\bf n}_3&=&-{\bf e}_8\vphantom{{1\over2}}
\quad\mbox{($\theta=0$, $\phi$, $\chi_1$, and $\chi_2$ arbitrary).}
\end{eqnarray}
\end{mathletters}
These unit vectors correspond to the following operator relations for
the projectors onto the original basis:
\begin{mathletters}
\begin{eqnarray}
|1\rangle\langle1|&=&
{1\over3}\Biggl[1+
\sqrt3\Biggl({\sqrt3\over2}\lambda_3+{1\over2}\lambda_8
\Biggr)\Biggr]\;,\\
|2\rangle\langle2|&=&
{1\over3}\Biggl[1+
\sqrt3\Biggl(-{\sqrt3\over2}\lambda_3+{1\over2}\lambda_8
\Biggr)\Biggr]\;,\\
|3\rangle\langle3|&=&
{1\over3}(1-\sqrt3\lambda_8)\vphantom{\Biggl(}\;.
\end{eqnarray}
For later use, we note here the operator expansions for the transition 
operators in the original basis:
\begin{eqnarray}
|1\rangle\langle2|&=&{1\over2}(\lambda_1+i\lambda_2)\;,\qquad
|2\rangle\langle1|={1\over2}(\lambda_1-i\lambda_2)\;,\\
|1\rangle\langle3|&=&{1\over2}(\lambda_4+i\lambda_5)\;,\qquad
|3\rangle\langle1|={1\over2}(\lambda_4-i\lambda_5)\;,\\
|2\rangle\langle3|&=&{1\over2}(\lambda_6+i\lambda_7)\;,\qquad
|3\rangle\langle2|={1\over2}(\lambda_6-i\lambda_7)\;.
\end{eqnarray}
\label{eq:oprelations}
\end{mathletters}

The geodesic distance between pure states, according to the unitarily
invariant Fubini-Studi metric \cite{Gibbons1992}, is given by the 
Hilbert-space angle between the states, i.e., 
$\cos s_{\rm FS}=|\langle\psi|\psi'\rangle|$.  For infinitesimally separated
pure states, $|\psi\rangle$ and $|\psi'\rangle=|\psi\rangle+|d\psi\rangle$, 
the distance gives the Fubini-Studi line element
\begin{equation}
ds_{\rm FS}^2=
1-|\langle\psi|\psi'\rangle|^2=
\langle d\psi|d\psi\rangle-|\langle\psi|d\psi\rangle|^2\;.
\end{equation}
We choose to rescale all lengths by a factor of $\sqrt3$, giving a line 
element
\begin{equation}
ds^2=3ds_{\rm FS}^2=d{\bf n}\cdot d{\bf n}\;,
\end{equation}
so that the metric on the submanifold of pure states is the metric induced 
by the natural metric on the unit sphere in eight dimensions.  In the 
co\"ordinates of Eq.~(\ref{eq:psi}), the rescaled line element becomes
\begin{eqnarray}
ds^2&=&3\Bigl(
d\theta^2+
\sin^2\!\theta\,d\phi^2+
\sin^2\!\theta\cos^2\!\phi(1-\sin^2\!\theta\cos^2\!\phi)\,d\chi_1^2
\nonumber\\
&\mbox{}&\phantom{3\Bigl(}
+\sin^2\!\theta\sin^2\!\phi(1-\sin^2\!\theta\sin^2\!\phi)\,d\chi_2^2
-2\sin^4\!\theta\sin^2\!\phi\cos^2\!\phi\,d\chi_1\,d\chi_2\Bigr)
\;.
\end{eqnarray}
The corresponding unitarily invariant volume element on the space of pure
states is 
\begin{equation}
d\Omega_{\bf n}=9\sin^3\!\theta\cos\theta\sin\phi\cos\phi\, 
d\theta\,d\phi\,d\chi_1\, d\chi_2
\;,
\end{equation}
which gives a total volume  
\begin{mathletters}
\begin{equation}
{\cal V}=\int d\Omega_{\bf n}={9\pi^2\over2}
\end{equation}
(notice that the volume relative to the original Fubini-Studi scaling is 
${\cal V}_{\rm FS}={\cal V}/9=\pi^2/2$).  We can also show that the
components of ${\bf n}$ satisfy 
\begin{eqnarray}
&\mbox{}&\int d\Omega_{\bf n}\,n_j=0\;,\\
&\mbox{}&\int d\Omega_{\bf n}\,n_j n_k=\frac{9\pi^2}{16}\delta_{jk}=
{{\cal V}\over8}\,\delta_{jk}\;.
\end{eqnarray}
\label{eq:integrals}
\end{mathletters}

\section{Mixtures of maximally mixed \\ and maximally entangled states}
\label{sec:epmax}

In this section we deal with two qutrits, labeled $A$ and $B$.  We
consider a class of two-qutrit states, specifically mixtures of the 
maximally mixed state, $M_9={1\over9}1\otimes1$, with a maximally 
entangled state, which we can choose to be
\begin{equation}
|\Psi\rangle={1\over\sqrt3}\Bigl(
|1\rangle\otimes|1\rangle
+|2\rangle\otimes|2\rangle
+|3\rangle\otimes|3\rangle\Bigr)
\;.
\label{eq:maxent}
\end{equation}
Such mixtures have the form
\begin{equation}
\rho_\epsilon=(1-\epsilon)M_9+\epsilon|\Psi\rangle\langle\Psi|
\;,
\label{eq:epmaxent}
\end{equation}
where $0\le\epsilon\le1$.  A state of the two qutrits is {\it separable\/} 
if it can be written as an ensemble of product states.  In this section we
show that the state~(\ref{eq:epmaxent}) is separable if and only if 
$\epsilon\le1/4$. 

In analogy to Eq.~(\ref{eq:rhooneexp}), any state $\rho$ of two qutrits can
be expanded uniquely as
\begin{equation}
\rho=\frac{1}{9}c_{\alpha\beta}\lambda_\alpha\otimes\lambda_\beta
\;,
\end{equation}
where the expansion coefficients are given by
\begin{equation}
c_{\alpha\beta}={9\over4}{\rm tr}(\rho\lambda_\alpha\otimes\lambda_\beta)
\;.
\label{eq:ctwo}
\end{equation}
Normalization requires that $c_{00}=3/2$.  Using the 
relations~(\ref{eq:oprelations}), we can find the operator expansion for 
the maximally entangled state~(\ref{eq:maxent}): 
\begin{eqnarray}
|\Psi\rangle\langle\Psi|&=&
\sum_{a,b}|a\rangle\langle b|\otimes|a\rangle\langle b|
\nonumber\\
&=& \frac{1}{9}
\Biggl(
1\otimes1+
\frac{3}{2}\Bigl(
\lambda_1\otimes\lambda_1
-\lambda_2\otimes\lambda_2
+\lambda_3\otimes\lambda_3
\nonumber\\
&\mbox{}&\hphantom{\frac{1}{9}\!\Biggl(1}
+\lambda_4\otimes\lambda_4
-\lambda_5\otimes\lambda_5
+\lambda_6\otimes\lambda_6
-\lambda_7\otimes\lambda_7
+\lambda_8\otimes\lambda_8
\Bigr)\Biggr)
\;.
\label{eq:opmaxent}
\end{eqnarray}
Hence the operator expansion for the mixed state~(\ref{eq:epmaxent}) is
\begin{eqnarray}
\rho_\epsilon&=&
\frac{1}{9}
\Biggl(
1\otimes1+
\frac{3\epsilon}{2}\Bigl(
\lambda_1\otimes\lambda_1
-\lambda_2\otimes\lambda_2
+\lambda_3\otimes\lambda_3
\nonumber\\
&\mbox{}&\hphantom{\frac{1}{9}\!\Biggl(1}
+\lambda_4\otimes\lambda_4
-\lambda_5\otimes\lambda_5
+\lambda_6\otimes\lambda_6
-\lambda_7\otimes\lambda_7
+\lambda_8\otimes\lambda_8
\Bigr)\Biggr)
\;,
\label{eq:opepmaxent}
\end{eqnarray}
from which we can read off the expansion coefficients~(\ref{eq:ctwo}) for 
the state $\rho_\epsilon$:
\begin{mathletters}
\begin{eqnarray}
&\mbox{}&c_{0j}=c_{j0}=0\;,\quad c_{jk}=0\quad\mbox{for $j\ne k$,}\\
&\mbox{}&c_{11}=-c_{22}=c_{33}=c_{44}=-c_{55}=c_{66}=-c_{77}=c_{88}=
{3\epsilon\over2}\;.
\label{eq:cjj}
\end{eqnarray}
\end{mathletters}

The product pure states for two qutrits, $P_{{\bf n}_A}\otimes P_{{\bf n}_B}$,
constitute an overcomplete operator basis.  Thus we can expand any 
two-qutrit density operator in terms of them:
\begin{equation}
\rho=
\int d\Omega_{{\bf n}_A}\,d\Omega_{{\bf n}_B}\,
w({\bf n}_A,{\bf n}_B)
P_{{\bf n}_A}\otimes P_{{\bf n}_B}\;.
\end{equation}
Because of the overcompleteness, the expansion function 
$w({\bf n}_A,{\bf n}_B)$ is not unique.  Notice that the expansion 
coefficients $c_{\alpha\beta}$ of Eq.~(\ref{eq:ctwo}) can be written 
as integrals over the expansion function,
\begin{equation} 
c_{\alpha\beta}=
3\int d\Omega_{{\bf n}_A}\,d\Omega_{{\bf n}_B}\,
w({\bf n}_A,{\bf n}_B)
(\tilde n_A)_\alpha (\tilde n_B)_\beta
\;,
\label{eq:ctwoint}
\end{equation}
where $\tilde n_0\equiv1/\sqrt2$ and $\tilde n_j\equiv n_j$.

A two-qutrit state $\rho$ is {\it separable\/} if there exists an expansion 
function that is everywhere nonnegative.  For a separable qutrit state,
the expansion function $w({\bf n}_A,{\bf n}_B)$ can be thought of as
a normalized classical probability distribution for the unit vectors 
${\bf n}_A$ and ${\bf n}_B$, and the integral for $c_{\alpha\beta}$ in 
Eq.~(\ref{eq:ctwoint}) can be interpreted as a classical expectation value 
over this distribution, i.e.,
\begin{equation}
c_{\alpha\beta}=3E[(\tilde n_A)_\alpha(\tilde n_B)_\beta]\;.
\label{eq:cE}
\end{equation}

If the state $\rho_\epsilon$ is separable, we have from Eqs.~(\ref{eq:cjj})
and (\ref{eq:cE}) that
\begin{equation}
{\epsilon\over2}=
{1\over3}|c_{jj}|=
\Bigr|E[(n_A)_j(n_B)_j]\Bigr|
\le
{1\over2}\Bigl(E[(n_A)_j^2\,]+E[(n_B)_j^2\,]\Bigr)
\;.
\end{equation}
Adding over the eight values of $j$ gives
\begin{equation}
4\epsilon\le 
{1\over2}\Bigl(E[{\bf n}_A\cdot{\bf n}_A]+E[{\bf n}_B\cdot{\bf n}_B]
\Bigr)
=1
\;.
\label{eq:necessity}
\end{equation}
We can conclude that if $\rho_\epsilon$ is separable, then $\epsilon\le1/4$.

To prove the converse, we need to construct a product ensemble for
$\epsilon=1/4$.  For each pair $a$, $b$, with $b>a$, define four pure 
product states
\begin{equation}
|\Phi_z^{(a,b)}\rangle\equiv
{1\over\sqrt2}\Bigl(|a\rangle+z|b\rangle\Bigr)
\otimes{1\over\sqrt2}\Bigl(|a\rangle+z^*|b\rangle\Bigr)
\;,
\end{equation}
where $z$ takes on the values $\pm1$ and $\pm i$, so that 
$0=\sum_z z=\sum_z z^2$ and $4=\sum_z|z|^2$.  It is easy to
show that an ensemble of the twelve states $|\Phi_z^{(a,b)}\rangle$, 
all states contributing with the same probability, gives the state 
$\rho_{\epsilon=1/4}$, i.e., 
\begin{equation}
{1\over12}\sum_{
{\scriptstyle{a,b}\atop\scriptstyle{b>a}}}
\sum_z|\Phi_z^{(a,b)}\rangle\langle\Phi_z^{(a,b)}|=
{3\over4}M_9+{1\over4}|\Psi\rangle\langle\Psi|\;,
\label{eq:ensemble}
\end{equation}
thus concluding the proof that $\rho_\epsilon$ is separable if and
only if $\epsilon=1/4$.

This result should be contrasted with that for the corresponding
qubit state, a Werner state \cite{Werner1989}, which is separable
if and only if $\epsilon\le1/3$ \cite{Bennett1996a}.  This indicates that 
maximally entangled states of two qutrits are more entangled than maximally
entangled states of two qubits.

The separability boundary obtained in this section is a special case of 
a general result obtained by Vidal and Tarrach \cite{Vidal1999}, who 
showed the following: for a pair of systems, one with dimension $d_1$ and 
the other with dimension $d_2$, a mixture,
$(1-\epsilon)M_{d_1d_2}+\epsilon|\Psi\rangle\langle\Psi|$, of the maximally
mixed state $M_{d_1d_2}$ with any pure state $|\Psi\rangle$ is separable 
if and only if $\epsilon\le(1+d_1d_2a_1a_2)^{-1}$, where $a_1^2$ and 
$a_2^2$ are the two largest eigenvalues of the marginal density operators 
obtained from $|\Psi\rangle\langle\Psi|$.  The reason for presenting
the much more limited result of this section is, first, that the 
proof of necessity, culminating in Eq.~(\ref{eq:necessity}), has a
nice physical interpretation in terms of the correlation coefficients 
$E[(n_A)_j(n_B)_j]$ and, second, that the ensemble~(\ref{eq:ensemble}) 
is different from the one used by Vidal and Tarrach.

\section{Separability of states near the maximally mixed state}
\label{sec:sep}

This section deals with $N$-qutrit states of the form
\begin{equation}
\rho_\epsilon=(1-\epsilon)M_{3^N}+\epsilon\rho_1\;,
\label{eq:rhoep}
\end{equation}
where $M_{3^N}\equiv1\otimes\cdots\otimes1/3^N$ is the maximally mixed
state for $N$ qutrits and $\rho_1$ is any $N$-qutrit density operator.
In this section we establish upper and lower bounds on the size of the 
neighborhood of separable states surrounding the maximally mixed state.  
In particular, we show, first, that for $\epsilon\le(1+3^{2N-1})^{-1}$, 
all states of the form~(\ref{eq:rhoep}) are separable and, second, that for 
$\epsilon>(1+3^{N/2})^{-1}$, there are states of the form~(\ref{eq:rhoep})
that are not separable.  The approach we take in this section follows
slavishly the corresponding qubit analysis presented by Braunstein 
{\it et al.}\ \cite{Braunstein1999}.
 
Since the product pure states form an overcomplete operator basis, any 
$N$-qutrit state can be expanded as 
\begin{equation}
\rho=\int
d\Omega_{{\bf n}_1}\cdots d\Omega_{{\bf n}_N}\,
w({\bf n}_1,\ldots,{\bf n}_N)
P_{{\bf n}_1}\otimes\cdots\otimes P_{{\bf n}_N}
\;.
\end{equation}
The expansion function $w({\bf n}_1,\ldots,{\bf n}_N)$ is not unique, because 
of the overcompleteness of the pure product states.  An $N$-qutrit state is 
separable if there exists an expansion function that is everywhere nonnegative.

We can find a particular expansion function in the following way.  Any
$N$-qutrit density operator can be expanded uniquely as
\begin{equation}
\rho={1\over3^N}c_{\alpha_1\ldots\alpha_N}
\lambda_{\alpha_1}\otimes\cdots\otimes\lambda_{\alpha_N}\;,
\label{eq:rhoc}
\end{equation}
with the expansion coefficients given by
\begin{equation}
c_{\alpha_1\ldots\alpha_N}=
\left({3\over2}\right)^{\!N}\!
{\rm tr}(\rho\lambda_{\alpha_1}\otimes\cdots\otimes\lambda_{\alpha_N})\;.
\end{equation}
Using Eqs.~(\ref{eq:integrals}), we can write
\begin{equation}
\lambda_\alpha={16\over3\sqrt3\pi^2}
\int d\Omega_{\bf n}\,\bar n_\alpha P_{\bf n}
\;,
\end{equation}
where the barred components are defined by $\bar n_0=1/4\sqrt2$ and
$\bar n_j=n_j$.  Substituting this expression for $\lambda_\alpha$ into 
Eq.~(\ref{eq:rhoc}), we find that one choice for the expansion function is 
\begin{eqnarray}
w_\rho({\bf n}_1,\ldots,{\bf n}_N)&=&
\left({16\over9\sqrt3\pi^2}\right)^{\!N}\!
c_{\alpha_1\ldots\alpha_N}
(\bar n_1)_{\alpha_1}\cdots(\bar n_N)_{\alpha_N}\nonumber\\
&=&\left({2\over9\pi^2}\right)^{\!N}\!
{\rm tr}\Bigl(\rho
(1+4\sqrt3\,{\bf n}_1\cdot\bbox{\lambda})
\otimes\cdots\otimes
(1+4\sqrt3\,{\bf n}_N\cdot\bbox{\lambda})
\Bigr)\;.
\label{eq:wrho}
\end{eqnarray}

Now consider the product operator in Eq.~(\ref{eq:wrho}).  The pure-state
density operator $P_{\bf n}={1\over3}(1+\sqrt3\,{\bf n}\cdot\bbox{\lambda})$ 
has eigenvalues 1, 0, and 0.  This implies that ${\bf n}\cdot\bbox{\lambda}$
has eigenvalues $2/\sqrt3$, $-1/\sqrt3$, and $-1/\sqrt3$ and that each
term in the product operator, $1+4\sqrt3\,{\bf n}\cdot\bbox{\lambda}$, has 
eigenvalues 9, -3, and -3.  Hence the smallest eigenvalue of the product
operator is $9^{N-1}(-3)=-3^{2N-1}$.  The result is a lower bound on the 
expansion function~(\ref{eq:wrho}):
\begin{equation}
w_\rho({\bf n}_1,\ldots,{\bf n}_N)\ge
\left({2\over9\pi^2}\right)^{\!N}\!\times
\pmatrix{\mbox{smallest eigenvalue of}\cr\mbox{product operator}}
=-\left({2\over9\pi^2}\right)^{\!N}\!3^{2N-1}
\;.
\end{equation}

We can use this lower bound to place a similar lower bound on the expansion
function for a state of the form~(\ref{eq:rhoep}).  Since the expansion
function for the maximally mixed state $M_{3^N}$ is the uniform distribution
$(2/9\pi^2)^N$, we have
\begin{equation}
w_{\rho_\epsilon}({\bf n}_1,\ldots,{\bf n}_N)=
(1-\epsilon)\left({2\over9\pi^2}\right)^{\!N}
+\epsilon w_{\rho_1}
\ge
\left({2\over9\pi^2}\right)^{\!N}\! 
\Bigl(1-\epsilon(1+3^{2N-1})\Bigr)
\;.
\end{equation}
We conclude that if
\begin{equation}
\epsilon\le{1\over1+3^{2N-1}}\;,
\end{equation}
$w_{\rho_\epsilon}$ is nonnegative, and thus the qutrit state 
$\rho_\epsilon$ of Eq.~(\ref{eq:rhoep}) is separable.  This establishes a 
lower bound on the size of the separable neighborhood surrounding the 
maximally mixed state.

We turn now to obtaining an upper bound on the size of the separable
neighborhood.  Consider two particles, each with spin $(3^{N/2}-1)/2$
($N$ even) and thus each having a $(d=3^{N/2})$-dimensional Hilbert space.
We can think of each particle as being an aggregate of $N/2$ spin-1
particles (qutrits).  We consider the following joint density operator for 
the two particles (or of the $N$ qutrits),
\begin{equation}
\rho_{\epsilon}=(1-\epsilon) M_{d^2}+\epsilon|\phi\rangle\langle\phi|
\;,
\label{eq:rhophi}
\end{equation}
where $M_{d^2}=1/d^2$ is the maximally mixed state in $d^2=3^N$ dimensions, 
and 
\begin{equation}
|\phi\rangle\equiv
{1\over\sqrt d}
\Bigl(
|1\rangle\otimes|1\rangle+|2\rangle\otimes|2\rangle+\cdots+
|d\rangle\otimes|d\rangle
\Bigr)
\label{eq:phi}
\end{equation}
is a maximally entangled state for the two particles.

Now project each particle onto the subspace spanned by $|1\rangle$,
$|2\rangle$, and $|3\rangle$.  For each particle this subspace can 
be thought of as the Hilbert space of a single qutrit; thus the projection 
leaves a joint state of two qutrits.  The projection is effected by 
the projection operator
\begin{equation}
\Pi=\sum_{a=1}^3|a\rangle\langle a|\otimes
\sum_{b=1}^3|b\rangle\langle b|
\;,
\end{equation}
which is the unit operator in the projected two-qutrit space, i.e.,
$\Pi=1\otimes1=9M_{9}$.  The normalized state after projection is
\begin{equation}
\tilde\rho_\epsilon={\Pi\rho_\epsilon\Pi\over
{\rm tr}(\rho_\epsilon\Pi)}
={1\over A}
\biggl((1-\epsilon){9\over d^2}M_9+
{3\epsilon\over d}|\Psi\rangle\langle\Psi|
\biggr)
=(1-\epsilon')M_9+\epsilon'|\Psi\rangle\langle\Psi|
\;,
\label{eq:projstate}
\end{equation}
where 
\begin{equation}
A={\rm tr}(\rho_\epsilon\Pi)=
(1-\epsilon){9\over d^2}+{3\epsilon\over d}=
{9\over d^2}\Bigl(1+\epsilon(d/3-1)\Bigr)
\end{equation}
is a normalization factor, $|\Psi\rangle$ is the maximally entangled
state of two qutrits given in Eq.~(\ref{eq:maxent}), and 
\begin{equation}
\epsilon'=
{3\epsilon/d\over A}=
{\epsilon d/3\over1+\epsilon(d/3-1)}
\;.
\end{equation}

The projected state $\tilde\rho_\epsilon$ is the state~(\ref{eq:epmaxent}) 
considered in Sec.~\ref{sec:epmax}, a mixture of the maximally mixed state 
for two qutrits, $M_9$, and the maximally entangled state $|\Psi\rangle$.  
As shown in Sec.~\ref{sec:epmax}, this state is nonseparable for 
$\epsilon'>1/4$, which is equivalent to 
\begin{equation}
\epsilon>{1\over1+d}={1\over1+3^{N/2}}\;.
\label{eq:upperbound}
\end{equation}
Moreover, since the local projections on the two particles cannot create 
entanglement from a separable state, we can conclude that the 
state~(\ref{eq:rhophi}) of $N$ qutrits is nonseparable under the same 
conditions.  This result establishes an upper bound, scaling like $3^{-N/2}$,
on the size of the separable neighborhood around the maximally mixed state.

\begin{acknowledgements}
This work, undertaken while GJM was a visitor in the Center for Advanced
Studies at the University of New Mexico, was supported in part by the Center
and by the Office of Naval Research (Grant No.~N00014-93-1-0116).
\end{acknowledgements}

\end{document}